\documentclass[pra,showpacs,superscriptaddress]{revtex4}

\def\var{\varepsilon}

\begin{document}

\title{Stabilizing quantum metastable states\\ in a time-periodic potential}
\author{Choon-Lin Ho}
\affiliation{Department of Physics, Tamkang University, Tamsui
25137, Taiwan}
\author{Chung-Chieh Lee}
\affiliation{Department of Physics, Tamkang University, Tamsui
25137, Taiwan} \affiliation{ Department of Electronic Engineering,
Tung Nan Institute of Technology, Shenkeng, Taipei 222, Taiwan}
\date{Aug 13, 2004}

\begin{abstract}
Metastability  of a particle trapped in a well with a
time-periodically oscillating barrier is studied in the Floquet
formalism. It is shown that the oscillating barrier causes the
system to decay faster in general. However, avoided crossings of
metastable states can occur with the less stable states crossing
over to the more stable ones.  If in the static well there exists
a bound state, then it is possible to stabilize a metastable state
by adiabatically increasing the oscillating frequency of the
barrier so that the unstable state will eventually cross over to
the stable bound state. It is also found that increasing the
amplitude of the oscillating field may change a direct crossing of
states into an avoided one.
\end{abstract}
\maketitle

\section{Introduction}

Ever since the advent of quantum mechanics, quantum tunneling has
been an important and fascinating subject. This phenomenon arises
frequently in physics.  In fact, one of the first successful
applications of quantum mechanics has been the explanation of the
$\alpha$ decay of atoms as a quantum tunneling process
\cite{alpha}.  Recent examples include tunneling phenomena in
semiconductors and superconductors \cite{SS}, in Josephson
junction systems \cite{JJS}, resonant tunneling in heterojunction
nanostructures \cite{Esaki}, tunneling ionization of atoms
\cite{DK}, photon-assisted tunneling in superconducting junctions
and semiconductor superlattices \cite{PAT}, etc.

In cosmology, quantum metastable states play an essential role in
some versions of the inflationary models of the early universe
\cite{Guth}. In these models inflation of the early universe is
governed by a Higgs field trapped in a metastable state. Inflation
ends when the metastable state decays to the true ground state of
the universe. During inflation the universe expands exponentially.
It is thus obvious that the metastable state of the Higgs field is
trapped in a rapidly varying potential.  The problem is therefore
a truly time-dependent one. Unfortunately, owing to the inherent
difficulties of the problem, more often than not one has to
consider the decay of the Higgs field in a quasistationary
approximation, in which the decay is studied assuming a static
potential \cite{Kolb}. Surely this approximation is hard to
justify, but for the present one has to be content with it.
Ultimately one hopes to be able to tackle the nonstationary case.
To this end, it is desirable to gain some insights first by
studying metastability in time-dependent potential in simple
quantum-mechanical models.

An early attempt at studying the effects of time-varying forces on
quantum metastability appears in Fisher's work \cite{Fisher},
which was motivated by an experiment on quantum tunneling of the
phase in a current-biased Josephson junction with a weak microwave
perturbation \cite{MDC}.  In this work  Fisher considered the
general problem of quantum tunneling in a metastable well with a
weak oscillatory force. There he reformulated the standard WKBJ
approach to quantum decay in order to include a weak
time-dependent perturbation.  For a class of metastable potentials
which interpolates between the cubic potential and a truncated
harmonic-oscillator potential, he showed that the decay rate is
generally enhanced by the weak oscillatory force.  The potential
considered by Fisher has a number of oscillator-like levels near
its minimum.  The opposite situation where only two levels are
present was considered by Sokolovski \cite{Sokolovski}, who
studied the effect of a small ac field mixing two levels in the
well on the tunneling rate in a semiclassical framework.

The results in \cite{Fisher,Sokolovski} are quite general for a
class of weak oscillatory forces.  However, it is desirable to
consider other possibilities, e.g., exact solutions and/or
nonperturbative results.  In this respect, we had considered
previously \cite{LeeHo} an exactly solvable quantum metastable
system with a moving potential which has height and width scaled
in a specific way introduced by Berry and Klein \cite{BK}.  We
found that in this model a small but finite nondecay probability
could persist at large time limit for an expanding potential.

In this paper we consider another simple driven quantum metastable
model in which a particle is trapped in a well with a periodically
driven rectangular barrier. In order to do away with any
restriction of amplitude or frequency of the driving force, and of
the number of states in the potential, we treat the problem in the
framework of the Floquet formalism \cite{Shirley,Sambe,LR,GH}.  An
exact expression determining the Floquet quasienergies of stable
or metastable states in the well is derived. From the solution of
this equation we find that while the oscillating barrier makes the
system decay faster in general, there is the possibility that
avoided crossings of metastable states can occur with the less
stable states crossing over to the more stable ones.

That an oscillating potential can affect the tunneling property of
a system has also been noticed before, e.g., in quantum transport
process \cite{BL,LR,GH,CDT}. In \cite{CDT,GH} it was found that a
particle can be localized in one side of a time-dependent double
well if the amplitude and the frequency of the oscillating field
were chosen properly.  In \cite{LR} it was demonstrated that a
propagating particle at appropriate incident energy can be trapped
into a bound state by an oscillating square well.  Our example
shows how a time-periodic field can modify the metastability of a
decaying state.

\section{The model}

The model we consider consists of a particle of mass $m$ trapped
in a square well with a harmonically oscillating barrier,
\begin{eqnarray}
 V(x,t)=\left\{
  \begin{array}{llll}
    \infty~, & x<0~, \\
    0~, & 0\leq x<a~,\\
    V_{0}+V_{1}\cos(\omega t)~, & a\leq x\leq b~, \\
    V_0^\prime ~, & x>b~.
  \end{array} \right.
  \label{potential}
\end{eqnarray}
Here $V_0,V_1, V_0^\prime$ and $\omega$ are positive parameters,
with $V_0^\prime <V_0$ and $V_1<V_0-V_0^\prime$. According to the
Floquet theorem, the wave function of a time-periodic system has
the form $\Psi_{\var}(x,t)=e^{-i\var t/\hbar}\Phi_{\var}(x,t)$,
where $\Phi_{\var}(x,t)$ is a periodic function with the period
$T=2\pi/\omega$, i.e., $\Phi_{\var}(x,t+T)=\Phi_{\var}(x,t)$, and
$\var$ is the Floquet quasienergy, which we will call Floquet
energy for brevity. It should be noted that the Floquet energy is
determined only modulo $\hbar\omega$. For if $\{\var, \Phi_\var\}$
is a solution of the Schr\"odinger equation, then
$\{\var^\prime=\var+n\hbar\omega~, \Phi_{\var^\prime}=\Phi_\var
\exp(in\omega t)\}$ is also a solution for any integer $n$.  But
they are physically equivalent as the total wave function
$\Psi_\var$ is the same \cite{Sambe}. All physically inequivalent
states can be characterized by their reduced Floquet energies in a
zone with a width $\hbar\omega$. We therefore consider solutions
of $\var$ only in the first Floquet zone, i.e., $\var\in
[0,\hbar\omega)$.

Following the procedures described in \cite{LR} (see also
\cite{BL}), we get the wave function as follows:
\begin{eqnarray}
 \Psi(x,t)&=& e^{-i\var t/\hbar}\Phi_\var (x,t)\\ \nonumber
 &=& e^{-i\var t/\hbar}\left\{
  \begin{array}{lll}
     \sum_{n=-\infty}^\infty A_{n}\sin(k_{n}x)
    e^{-in\omega t}~, & 0\leq x<a~, \\
   \sum_{n=-\infty}^\infty\sum_{l=-\infty}^\infty
    \left(a_{l}e^{q_{l}x}+b_{l}e^{-q_{l}x}\right)
 J_{n-l}\left(V_1/\hbar\omega\right)e^{-in\omega t}~, & a\leq x\leq b~,
\\    \sum_{n=-\infty}^\infty t_{n}e^{ik'_{n}x}e^{-in\omega t}~, &
x>b~,
  \end{array} \right.
  \label{wf}
\end{eqnarray}
where
\begin{eqnarray}
 k_{n}&=&\sqrt{2m(\var+n\hbar\omega)}/\hbar~, \nonumber\\
q_{l}&=&\sqrt{2m(V_0-\var-l\hbar\omega)}/\hbar~,\label{k}\\
k'_{n}&=&\sqrt{2m(\var+n\hbar\omega-V'_{0})}/\hbar~, \nonumber
\end{eqnarray}
 and $J_n$ is
the Bessel function.  In the region $x>b$, we have adopted Gamow's
outgoing boundary condition, namely, there is no particle
approaching the barrier from the right \cite{alpha}. Matching the
wave function and its first derivative at the boundaries $x=a$ and
$x=b$, we obtain the relations among the coefficients
$A_{n},a_{n},b_{n}$ and $t_{n}$:
\begin{eqnarray}
 A_{n}\sin(k_{n}a) &=&
 \sum_{l}\left(a_{l}e^{q_{l}a}+b_{l}e^{-q_{l}a}\right)J_{n-l}
 (\alpha)~,
 \nonumber
 \\
 k_{n}A_{n}\cos(k_{n}a) &=&
 \sum_{l}q_{l}\left(a_{l}e^{q_{l}a}-b_{l}e^{-q_{l}a}\right)
 J_{n-l}(\alpha)~,
 \label{b.c.1}
 \\
 t_{n}e^{ik'_{n}b} &=&
 \sum_{l}\left(a_{l}e^{q_{l}b}+b_{l}e^{-q_{l}b}\right)J_{n-l}
 (\alpha)~,
 \nonumber
 \\
 ik'_{n}t_{n}e^{ik'_{n}b} &=&
 \sum_{l}q_{l}\left(a_{l}e^{q_{l}b}-b_{l}e^{-q_{l}b}\right)
 J_{n-l}(\alpha)~,
 \nonumber
\end{eqnarray}
where $\alpha\equiv V_{1}/\hbar\omega$. The Floquet energy is
determined from these relations by demanding nontrivial solutions
of the coefficients.  In practice, however, we must truncate the
above equations to a finite number of terms, or sidebands as they
are usually called in the literature, eg., $n=0, \pm1,\ldots,\pm
N$ . The number $N$ is determined by the frequency and the
strength of the oscillation as $N>V_{1}/\hbar\omega$ \cite{LR}.

We proceed to determine the
 Floquet energy as follows.
We first separate the boundary conditions for the central band
($n=0$) from those for the subbands ($n\neq 0$) in
Eq.(\ref{b.c.1}).  From the boundary conditions for the subbands
$(n\neq 0)$, one can relate the coefficients $a_{l}$ and $b_{l}$
($l\neq 0$) with the coefficients $a_{0}$ and $b_{0}$ through the
following two relations:
\begin{eqnarray}
 a_{l} &=& f_{la}(k_{0},k'_{0},\omega,V_{1})a_{0}+f_{lb}(k_{0},
 k'_{0},\omega,V_{1})b_{0}~,
 \label{a_l}
 \\
 b_{l} &=& g_{la}(k_{0},k'_{0},\omega,V_{1})a_{0}+g_{lb}(k_{0},
 k'_{0},\omega,V_{1})b_{0}~,
 \label{b_l}
\end{eqnarray}
where $f$'s and $g$'s are functions determined as follows.
Eliminating the $A_n$'s and $t_n$'s in Eq.(\ref{b.c.1}), we can
obtain
\begin{eqnarray}
 &&A^{-}_{n,n}e^{q_{n}a}J_{0}a_{n} +A^{+}_{n,n}e^{-q_{n}a}J_{0}b_{n}
 +\sum_{l\neq n,0}A^{-}_{n,l}e^{q_{l}a}J_{n-l}a_{l}
 +\sum_{l\neq n,0}A^{+}_{n,l}e^{-q_{l}a}J_{n-l}b_{l}
 \nonumber \bigskip \\
 &=&
 -A^{-}_{n,0}e^{q_{0}a}J_{n}a_{0}-A^{+}_{n,0}
 e^{-q_{0}a}J_{n}b_{0}~,
 \label{b.c.neq0_1}
\end{eqnarray}
and
\begin{eqnarray}
 &&B^{+}_{n,n}e^{q_{n}b}J_{0}a_{n}+B^{-}_{n,n}
 e^{-q_{n}b}J_{0}b_{n}
 +\sum_{l\neq n,0}B^{+}_{n,l}e^{q_{l}b}J_{n-l}a_{l}
 +\sum_{l\neq n,0}B^{-}_{n,l}e^{-q_{l}b}J_{n-l}b_{l}
 \nonumber \bigskip \\
 &=& -B^{+}_{n,0}e^{q_{0}b}J_{n}a_{0}-B^{-}_{n,0}
 e^{-q_{0}b}J_{n}b_{0}~,
 \label{b.c.neq0_2}
\end{eqnarray}
with
\begin{eqnarray}
 A^{\pm}_{n,l} \equiv \cos k_{n}a \pm \frac{q_{l}}{k_{n}}\sin
 k_{n}a~,~~{\rm  and}~~
 B^{\pm}_{n,l} \equiv 1 \pm i\frac{q_{l}}{k'_{n}}~.
\end{eqnarray}
Equations (\ref{b.c.neq0_1}) and (\ref{b.c.neq0_2}) allow us to
solve for $a_l$ and $b_l$ in terms of $a_0$ and $b_0$ in the forms
of Eqs.(\ref{a_l}) and (\ref{b_l}) by means of the Cramer's rule
in matrix algebra.  As mentioned before, in practice a truncated
version of Eqs. (\ref{b.c.neq0_1}) and (\ref{b.c.neq0_2}) has to
be used.

Using Eqs.(\ref{a_l}) and (\ref{b_l}) we can rewrite the boundary
conditions for the central band $n = 0$ as
\begin{eqnarray}
 A_0\sin(k_{0}a) &=&
 F_{1}(k_{0},k'_{0};\omega,V_{1})e^{q_{0}a}a_{0}+F_{2}(k_{0},k'_{0};
 \omega,V_{1})e^{-q_{0}a}b_{0}~,
 \nonumber
 \\
 k_{0}A_0\cos(k_{0}a) &=&
 F_{3}(k_{0},k'_{0};\omega,V_{1})q_{0}e^{q_{0}a}a_{0}-F_{4}
 (k_{0},k'_{0};
 \omega,V_{1})q_{0}e^{-q_{0}a}b_{0}~,
 \label{b.c.2}
 \\
 t_{0}e^{ik'_{0}b} &=&
 F_{5}(k_{0},k'_{0};\omega,V_{1})e^{q_{0}b}a_{0}+F_{6}
 (k_{0},k'_{0};\omega,V_{1})e^{-q_{0}b}b_{0}~, \nonumber
 \\
 ik'_{0}t_{0}e^{ik'_{0}b} &=&
 F_{7}(k_{0},k'_{0};\omega,V_{1})q_{0}e^{q_{0}b}a_{0}-F_{8}
 (k_{0},k'_{0};\omega,V_{1})q_{0}e^{-q_{0}b}b_{0}~, \nonumber
\end{eqnarray}
where the coefficients $F_{i}(k_{0};\omega,V_{1})\ (i=1,\dots,8)$
are
\begin{eqnarray*}
 F_{1}(k_{0},k'_{0};\omega,V_{1}) &=&
 J_{0}(\alpha)+e^{-q_{0}a}\sum_{l\neq 0}
 \left(f_{la}e^{q_{l}a}+g_{la}e^{-q_{l}a}\right)J_{-l}(\alpha)~,
 \\
 F_{2}(k_{0},k'_{0};\omega,V_{1}) &=&
 J_{0}(\alpha)+e^{q_{0}a}\sum_{l\neq 0}
 \left(f_{lb}e^{q_{l}a}+g_{lb}e^{-q_{l}a}\right)J_{-l}(\alpha)~,
 \\
 F_{3}(k_{0},k'_{0};\omega,V_{1}) &=&
 J_{0}(\alpha)+e^{-q_{0}a}\sum_{l\neq 0}
 \frac{q_{l}}{q_{0}}\left(f_{la}e^{q_{l}a}-g_{la}e^{-q_{l}a}\right)
 J_{-l}(\alpha)~,
 \\
 F_{4}(k_{0},k'_{0};\omega,V_{1}) &=&
 J_{0}(\alpha)-e^{q_{0}a}\sum_{l\neq 0}
 \frac{q_{l}}{q_{0}}\left(f_{lb}e^{q_{l}a}-g_{lb}e^{-q_{l}a}\right)
 J_{-l}(\alpha)~,
 \\
 F_{5}(k_{0},k'_{0};\omega,V_{1}) &=&
 J_{0}(\alpha)+e^{-q_{0}b}\sum_{l\neq 0}
 \left(f_{la}e^{q_{l}b}+g_{la}e^{-q_{l}b}\right)J_{-l}(\alpha)~,
 \\
 F_{6}(k_{0},k'_{0};\omega,V_{1}) &=&
 J_{0}(\alpha)+e^{q_{0}b}\sum_{l\neq 0}
 \left(f_{lb}e^{q_{l}b}+g_{lb}e^{-q_{l}b}\right)J_{-l}(\alpha)~,
 \\
 F_{7}(k_{0},k'_{0};\omega,V_{1}) &=&
 J_{0}(\alpha)+e^{-q_{0}b}\sum_{l\neq 0}
 \frac{q_{l}}{q_{0}}\left(f_{la}e^{q_{l}b}-g_{la}e^{-q_{l}b}\right)
 J_{-l}(\alpha)~,
 \\
 F_{8}(k_{0},k'_{0};\omega,V_{1}) &=&
 J_{0}(\alpha)-e^{q_{0}b}\sum_{l\neq 0}
 \frac{q_{l}}{q_{0}}\left(f_{lb}e^{q_{l}b}-g_{lb}e^{-q_{l}b}\right)
 J_{-l}(\alpha)~.
\end{eqnarray*}
By demanding nontrivial solutions of the coefficients $a_{0}$,
$b_{0}$, $A_0$, and $t_{0}$ in  Eq.(\ref{b.c.2}), we obtain an
equation which determines the Floquet energy $\var$:
\begin{eqnarray}
 F_{4}\frac{q_{0}}{k_{0}}\tan k_{0}a +F_{2}
 =\frac{F_{8}q_{0}+iF_{6}k'_{0}}{F_{7}q_{0}-iF_{5}k'_{0}}
 \left(F_{3}\frac{q_{0}}{k_{0}}\tan k_{0}a -F_{1}\right)
 e^{-2q_{0}(b-a)}~.
 \label{solution}
\end{eqnarray}
We recall here that $k_0, q_0, k'_0$ are functions of the Floquet
energy $\var$ [c.f. Eq.~(\ref{k})]. If the solutions $\var$ of
Eq.(\ref{solution}) are complex (real) numbers, the corresponding
Floquet states are metastable (stable) states. The nondecay
probability $P(t)$, which is the probability of the particle still
being trapped by the potential barrier at time $t>0$, is given by
\begin{eqnarray}
 P(t) &=& \frac{\int_{0}^{b}\left|\Psi(x,t)\right|^2dx}
 {\int_{0}^{b}\left|\Psi(x,0)\right|^2dx}
 \nonumber \\
 &=& e^{2Im(\var)t/\hbar} \frac{\int_{0}^{b}\left|\Phi_\var(x,t)\right|^2dx}
 {\int_{0}^{b}\left|\Phi_\var(x,0)\right|^2dx}
 \label{P(t)}\\
 &\equiv & e^{2Im(\var)t/\hbar} h(t)~,\nonumber
\end{eqnarray}
with $P(0)=1$. The imaginary part of the Floquet energy, which
enters $P(t)$ via the factor $\exp[2Im(\var)t/\hbar]$, gives a
measure of the stability of the system.  Unlike the static case,
however, here $P(t)$ is not a monotonic function of time, owing to
the time-dependent function $h(t)$ after the exponential factor in
Eq.({\ref{P(t)}). But since $h(t)$ is only a periodic function
oscillating between two values which are of order one, the
essential behavior of $P(t)$ at large times is still mainly
governed by the exponential factor.  Hence, as a useful measure of
the nondecay rate of the particle in the well, we propose a
coarse-grained nondecay probability $\bar{P}(t)$ defined as
\begin{eqnarray}
\bar{P}(t)\equiv  e^{2Im(\var)t/\hbar} \langle h(t) \rangle~,
\label{bP(t)}
\end{eqnarray}
where $\langle h(t) \rangle$ is the time average of $h(t)$ over
one period of oscillation. The graphs of $P(t)$ and $\bar{P}(t)$
will be given in the next section.

It is easily seen that the coefficients
$F_{i}(k_{0};\omega,V_{1})$ all approach one in the limit
$\alpha=V_1/\hbar\omega\to 0$,
\begin{eqnarray}
 \lim_{V_{1}/\omega\rightarrow 0}F_{i}(k_{0};\omega,V_{1}) \longrightarrow 1~,
 \ i=1,\dots,8~.
\end{eqnarray}
Hence in the limit $V_{1}\to 0$ or $\omega\to \infty$,
Eq.(\ref{solution}) reduces to the corresponding equation for the
case of static potential with potential $V_0$ in the region $a\leq
x \leq b$, and the Floquet energy in this limit is just the (real
or complex) eigenenergy of the static case. This is
understandable, since in the limit $V_1\to 0$ the potential
becomes static, and at high frequencies the particle in the well
will only see a time-averaged barrier of effective height $V_0$
\cite{BL}.

\section{Numerical results}

We now study numerical solutions of Eq.(\ref{solution}) with a
specific potential.  We take $a=1,~b=2,~V_0=15$, and
$V_0^\prime=V_0/2$ in the atomic units (a.u.) ($e=m_e=\hbar=1$).
In the static case this potential supports one bound state, with
energy $E_0/V_0=0.232123$,  and one metastable state, with complex
energy $E_1/V_0=0.864945-0.00255261i$.  For the oscillating
potential, we solve Eq.(\ref{solution}) in 2-sideband
approximation, i.e., we take $N=2$.  This is accurate enough for
oscillating frequency $\omega\geq V_1/2$.

Figures 1 and 2 present the graphs of the real and imaginary parts
of the Floquet energy ($\var/V_0$) as a function of
$\omega/V_0\geq 0.2$ with $V_1=0.1V_0$ and $0.2V_0$, respectively.
We find that the solutions of Eq.(\ref{solution}) have the form
$\var=\var_0 + n\omega$ ($n=0,\pm 1, \pm 2,\ldots$), with
$Re(\var_0)$ (the horizontal branch) laying close to the energies
$E_0$ and $Re(E_1)$ in the static potential. That is, these
branches of $Re(\var)$ emanate from either $E_0$ or $Re (E_1)$ at
$\omega=0$. Branches emerging from the same point have the same
imaginary part. Numerical results show that, with the barrier
oscillating, the stable state ($E_0$) in the static case becomes
unstable, and the unstable state ($E_1$) will decay even faster.
For simplicity, in Figs.~(1a) and (2a) we show only six branches
($n=0,\pm 1, \pm 2$ and $-3$) emerging from $Re (E_1)$, and only
the central branch ($n=0$) and a subband ($n=-1$) from $E_0$.  As
mentioned before, we only take solutions in the first Floquet
zone, $Re(\var)$ (modulo $\omega$), which are points under the
line $Re(\var)=\omega$.

In Fig.~3 we give the graphs of the probability density $|\Psi|^2$
in the well and the barrier with the same parameters as in Fig.~2
for the two metastable states at frequency $\omega/V_0=0.62$. The
Floquet energies of the less stable and the more stable state are
$\var/V_0=0.251714-0.004995i$ and $0.227343-0.001456i$,
respectively. Four time frames, namely, $\tau\equiv t\times V_0
=0,100, 200$ and $300~ a.u.$, are shown, with the probability
density normalized to unity within $0\leq x\leq 2$ at $\tau=0$.
One can clearly see that the less stable state (dashed curve)
decays much faster than the more stable state (solid curve). The
nondecay probability (oscillatory curve) $P(t)$, Eq.(\ref{P(t)}),
and the coarse-grained nondecay probability (monotonic curve)
$\bar{P} (t)$, Eq.(\ref{bP(t)}), of these two states are shown in
Fig.~4. It is clear that the coarse-grained function $\bar{P} (t)$
is a monotonic function, and does give a smooth measure of the
stability of the system.

From Figs.~1 and 2 we also see that a direct crossing occurs at
frequency $\omega\approx Re(E_1 - E_0)/2$ (point $c$). However, as
$\omega$ approaches the frequency $\omega\approx Re(E_1 -
E_0)=0.632822V_0$, an avoided crossing ($e,e^\prime$) between the
real parts of the Floquet energies occurs. Figure 2 indicates that
larger values of $V_1$ only enhance the instability of the system
and the repulsion between the two levels at avoided crossing. Thus
as the frequency $\omega$ is increased, the state emanating from
$E_1$ has Floquet energy with real part given by values along the
path $abb^\prime cdd^\prime e^\prime f$ (the dark dotted curve),
while the real part of Floquet energy of the state emerging from
$E_0$ lies along the path $cegg^\prime h$ (the solid curve).  The
imaginary parts of these two paths are depicted in Figs.~(1b) and
(2b).  One sees that an exchange of the imaginary parts takes
place at the avoided crossing $ee^\prime$. In Figs.~5 and 6, we
show the probability density of the two states within the
potential barrier just before and just after the avoided crossing.
Together with Fig.~3, these plots demonstrate clearly the
switching of the states.  Beyond the avoided crossing, the upper
state becomes more stable than the lower state.  This gives the
possibility of stabilizing an unstable state by an oscillating
field.  We recall that as $\omega\to \infty$, Eq.(\ref{solution})
reduces to the one for the static potential. In the example
considered here, the lower state supported by the static well is a
stable bound state, and hence the unstable upper Floquet state can
be made stable in the high frequency limit.  Even more simply, the
same aim can be achieved by adiabatically tuning down the
amplitude $V_1$ just after the avoided crossing, as in this limit
the potential becomes the static one.

Finally, it is interesting to note here the role of amplitude
$V_1$ of the oscillating barrier in the model. As we have seen,
the presence of a nonvanishing $V_1$ always makes the system less
stable.  However, if $V_1$ is reduced, an avoided crossing may
turn into a direct one. In the present case, the avoided crossing
$ee'$ changes into a direct crossing for $V_1/V_0<0.03$.
Conversely, increasing $V_1$ could change a direct crossing into
an avoided crossing.  At an avoided crossing, the imaginary parts
of the Floquet energies cross, while the real parts do not. At a
direct crossing, it is the real parts, not the imaginary parts,
that cross. But the more (less) stable state has the tendency to
become less (more) stable.  This is evident from the Floquet
energy at the direct crossing point $c$ in Figs.~1 and 2.  These
observations are consistent with the semiclassical results
obtained in \cite{Sokolovski} by perturbative methods.  Hence, by
a combination of adiabatic changes of the frequency and the
amplitude of the oscillating barrier, one can manipulate the
stability of different states in a quantum potential: tune up
$V_1$ until a direct crossing becomes an avoided one, increase
$\omega$ so that the avoided crossing is passed, then reduce $V_1$
to make the potential static. In the process, two states in the
well are interchanged.

\section{summary}

To summarize, our results show that an oscillating potential
barrier generally makes a metastable system decay faster. However,
the existence of avoided crossings of metastable states can switch
a less stable state to a more stable one.  If in the static well
there exists a bound state, then it is possible to stabilize a
metastable state by adiabatically changing the oscillating
frequency and amplitude of the barrier so that the unstable state
will eventually cross over to the stable bound state.  Thus a
time-dependent potential can be used to control the stability of a
particle trapped in a well.

Finally, we would like to comment on the differences between the
stabilization of the decaying state discussed in this work and the
interesting phenomenon of the suppression of ionization of atom
(also called stabilization of atom) in superintense, high
frequency laser pulses \cite{Sup-ion}. In this later phenomenon,
it was found that while an atom is generally ionized by absorption
of photons, the ionization rate of the atom can be dramatically
suppressed as the intensity (amplitude) of the laser field exceeds
a certain threshold.  Thus stabilization of the atom is attained
by increasing the intensity. On the contrary, stabilization of the
decaying state described here is achieved by increasing the
frequency of the oscillating barrier to the threshold at which an
avoided crossing in the Floquet energies takes place, regardless
of the amplitude of the field.  Also, in our case stabilization is
against quantum tunneling through a potential barrier, while in
the case of atomic stabilization it is against ionization by
photon absorption. Another difference of the two phenomena is that
an atom in the ground state is stable when not being irradiated by
the laser field, but the decaying state to be stabilized in our
system is already unstable even in the absence of the oscillating
field owing to quantum tunneling effect. Furthermore, suppression
of ionization can be studied using the methods of classical
nonlinear dynamics and chaos \cite{Classical}, but tunneling
through a barrier considered here is a characteristic and
fundamental quantum phenomenon.

\begin{acknowledgments}
This work was supported in part by the National Science Council of
the Republic of China through Grant No. NSC 92-2112-M-032-015 and
NSC 93-2112-M-032-009.
\end{acknowledgments}

\vskip 3truecm
\centerline{\bf Figure Captions}
\begin{description}
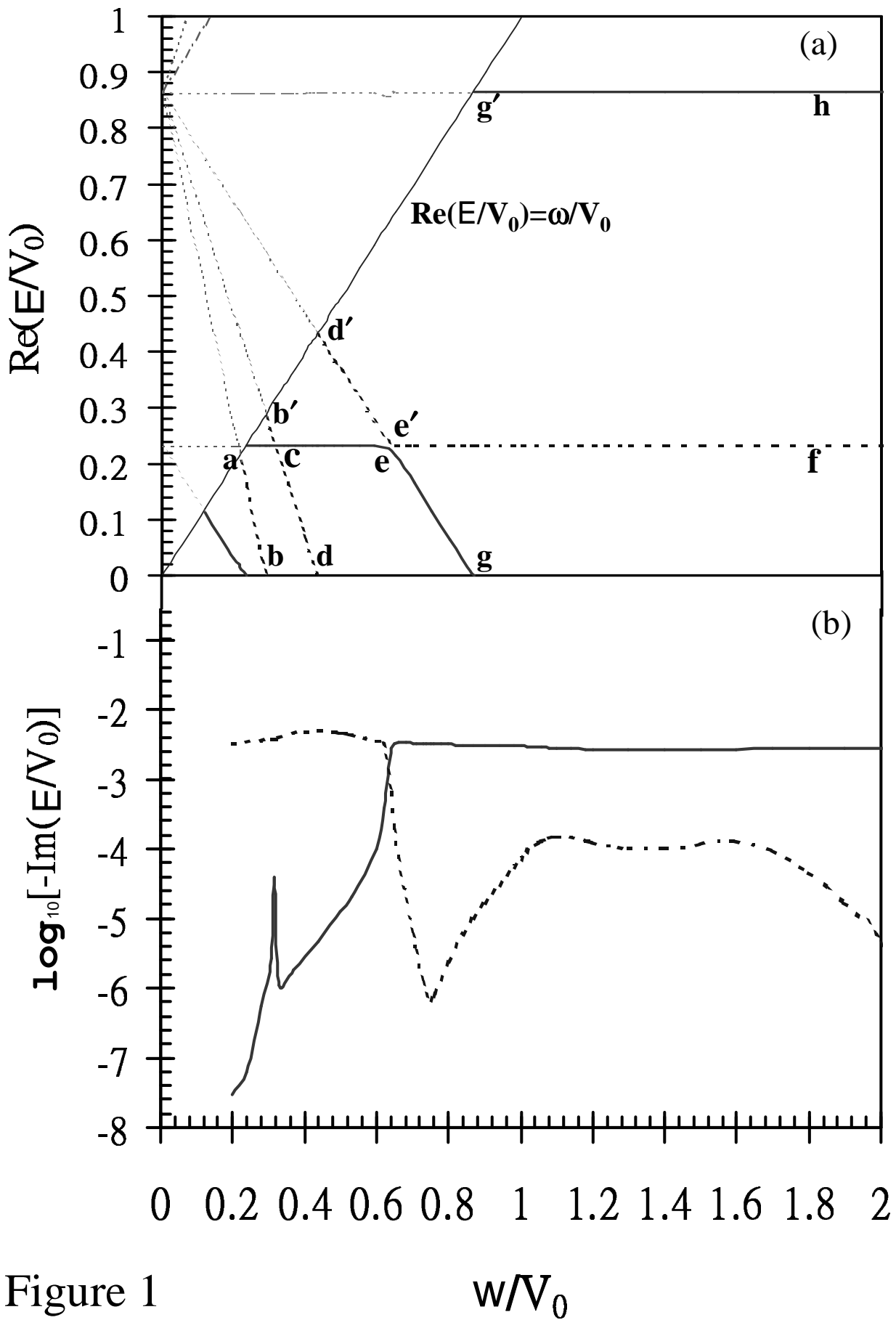
\item[Figure 1.]  The Floquet
energies ($\var/V_0$) of the two metastable states versus the
barrier oscillating frequency ($\omega/V_0$) for $V_0=15 a.u.,
V_0^\prime=V_0/2$ and $V_1/V_0=0.1$ in the atomic units (a.u.)
($e=m_e=\hbar=1$) . In (a) the real parts of the Floquet energies
are shown in the first Floquet zone under the line
$Re(\var)=\omega$ (the straight line). The light dotted lines show
how the different branches emanate from the two states in the
static case [with $E_0/V_0=0.232123$ and $Re(E_1)/V_0=0.864945$].
In (b) the corresponding imaginary parts of the Floquet energies
of the two states are plotted. The dotted curve corresponds to the
state with real parts given along the path $abb^\prime cdd^\prime
e^\prime f$, and the solid curve corresponds to the state with
real parts given along $cegg^\prime h$.

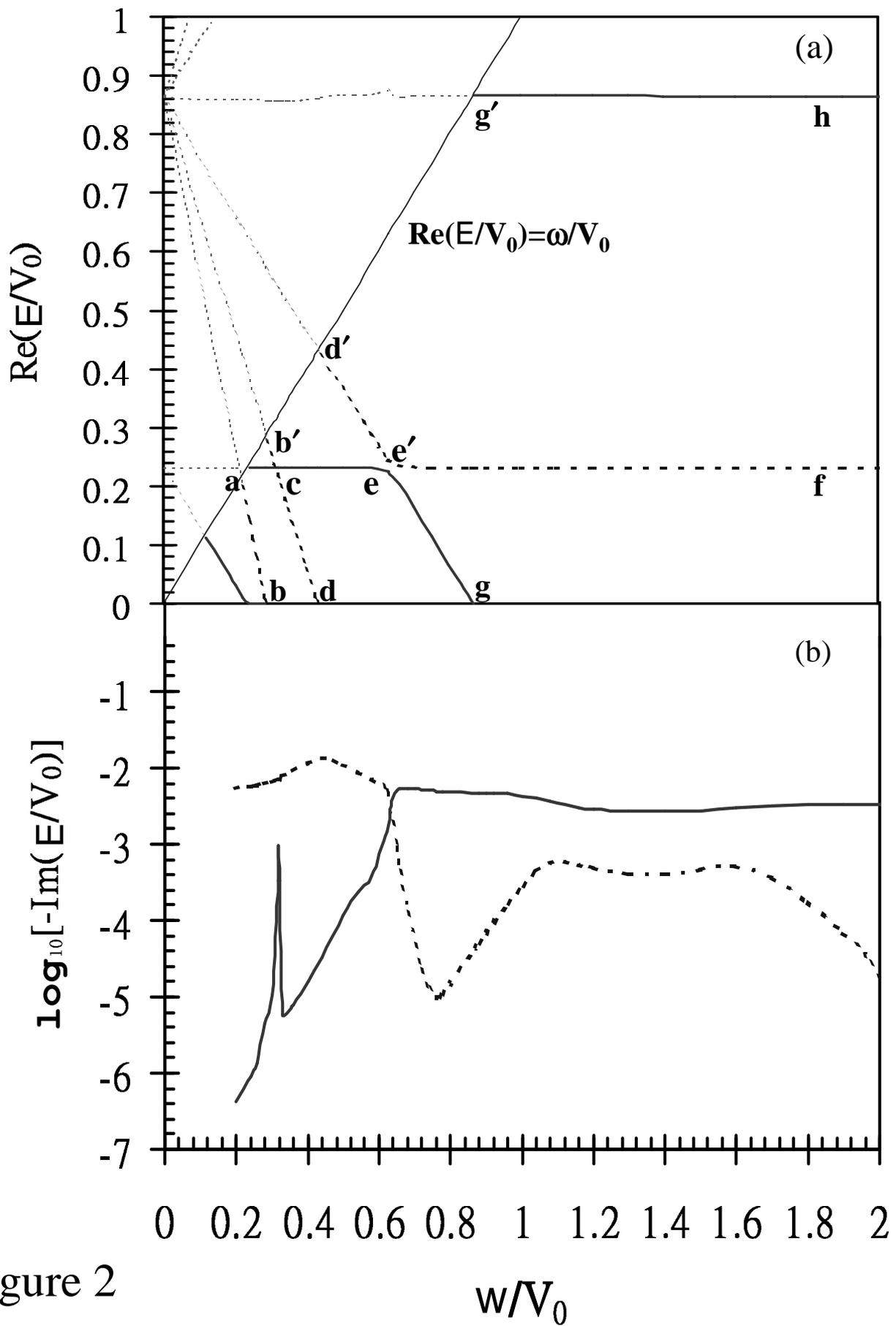
\item[Figure 2.]  Same plot as Fig.~1 for $V_0=15 a.u.,
V_0^\prime=V_0/2$, and $V_1/V_0=0.2$.

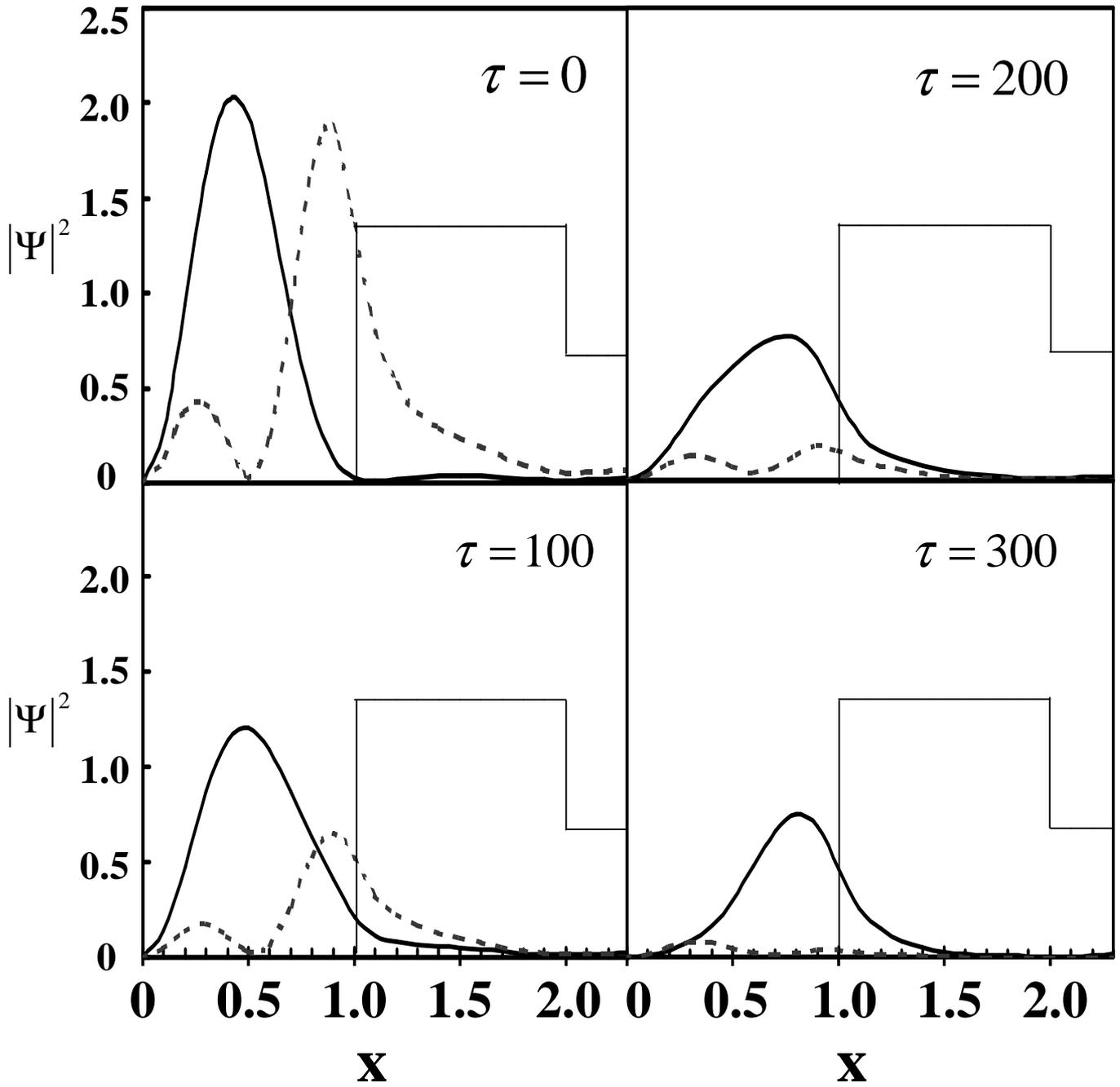
\item[Figure 3.]  Probability density $|\Psi|^2$
in the well and the barrier with the same parameters as in Fig.~2
for the two metastable states at frequency $\omega/V_0=0.62$.
Probability density normalized to unity within $0\leq x\leq 2$ at
$\tau\equiv t V_0 = 0 ~a.u.$.  The more (less) stable state is
indicated by solid (dashed) curve.

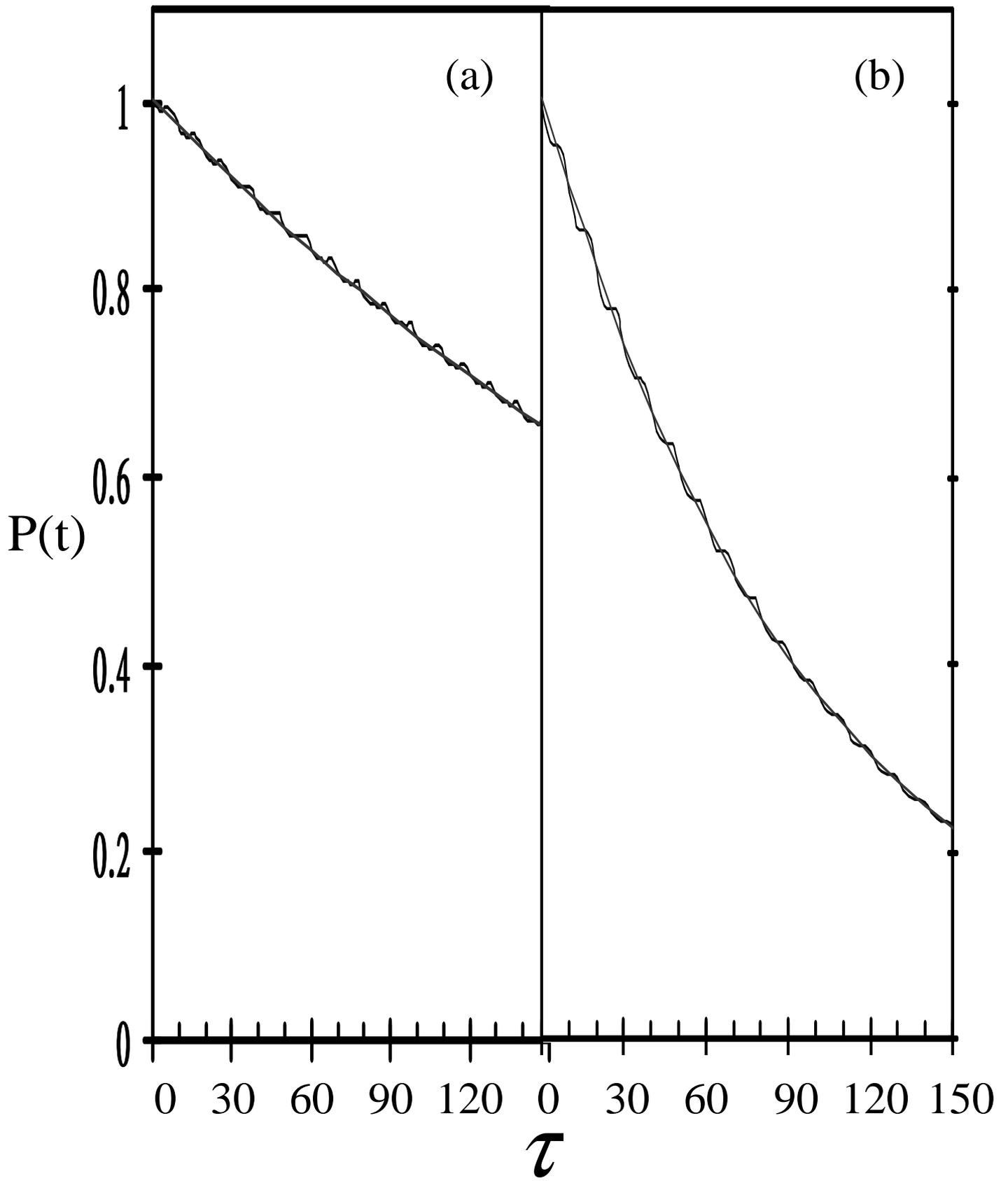
\item[Figure 4.] Nondecay probability (oscillatory curve)
$P(t)$ and the coarse-grained nondecay probability (monotonic
curve) $\bar{P} (t)$ as a function of time for the more stable (a)
and the less stable (b) state in Fig.~3.

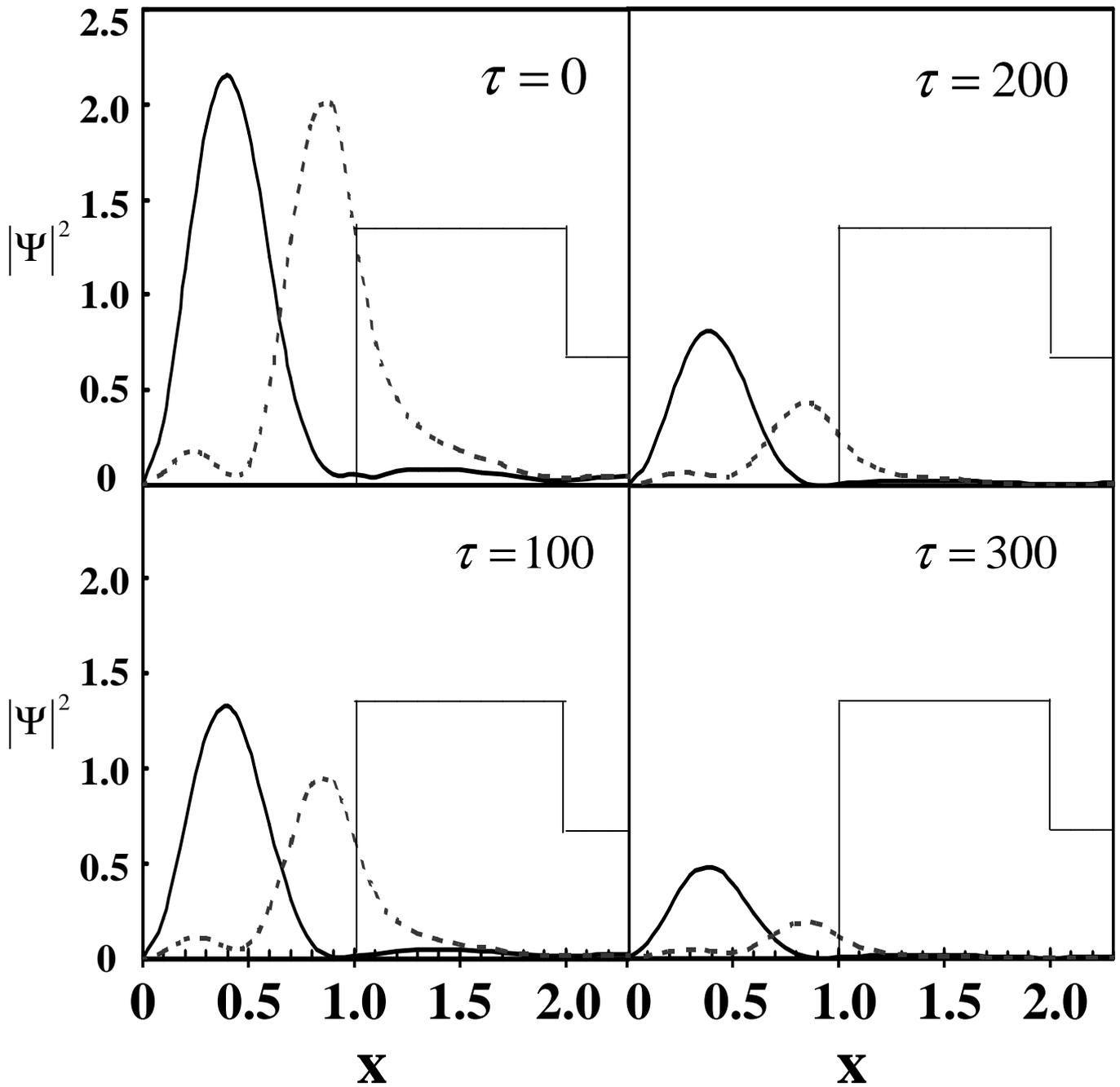
\item[Figure 5.]   Same plot as Fig.~3 at frequency $\omega/V_0=0.63$, just
before the avoided crossing. The lifetimes of these two metastable
states are comparable.

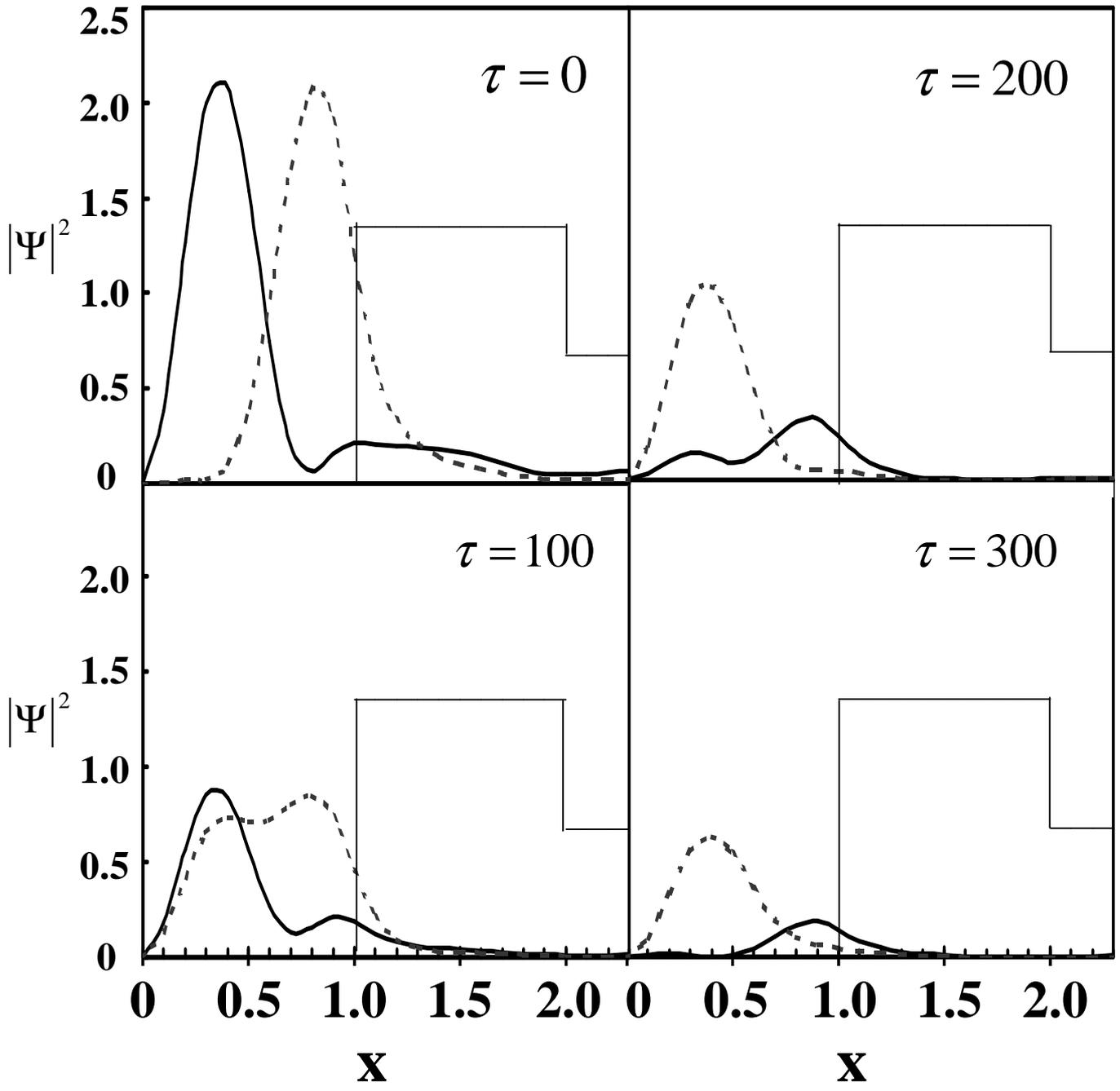
\item[Figure 6.]   Same plot as Fig.~3 at frequency $\omega/V_0=0.64$,
just after the avoided crossing. The dashed (solid) curve
represents the originally less (more) stable state, which now
becomes more (less) stable.

\end{description}

\end{document}